\documentstyle[multicol,aps,pre]{revtex}
\begin{document}
\title{Velocity Distribution of Topological Defects in Phase-Ordering 
Systems}
\author{A. J. Bray} 
\address{Department of Physics and Astronomy, The University, Manchester 
M13 9PL, UK. 
}
\date{\today}
\maketitle

\begin{abstract}
The distribution of interface (domain-wall) velocities ${\bf v}$ in 
a phase-ordering system is considered. Heuristic scaling arguments 
based on the disappearance of small domains lead to a power-law tail, 
$P_v(v) \sim v^{-p}$ for large $v$, in the distribution of 
$v \equiv |{\bf v}|$. The exponent $p$ is given by $p = 2+d/(z-1)$, 
where $d$ is the space dimension and $1/z$ is the growth exponent, 
i.e.\ $z=2$ for nonconserved (model A) dynamics and $z=3$ for the 
conserved case (model B). The nonconserved result is exemplified by 
an approximate calculation of the full distribution using a gaussian 
closure scheme. The heuristic arguments are readily generalized to 
systems described by a vector order parameter.  

\end{abstract}

\begin{multicols}{2}
\section{Introduction}
The theory of phase-ordering dynamics has seen significant advances in 
recent years \cite{Review}. Coarsening proceeds by the elimination 
of topological defects, e.g.\ domain walls, vortices, or strings, and 
various properties of the defect distribution have been investigated in 
some detail \cite{LM92,MG}.

In a recent paper, Mazenko \cite{Mazenko} has carried out the first 
investigation of the distribution of defect velocities in a phase-ordering 
system. Using an approximate `gaussian closure' scheme, he has computed 
the velocity distribution for point defects in $O(n)$ models with 
nonconserved order parameter. He finds the interesting result that the 
velocity distribution has a power-law tail.

In the present work we shall show that such power-law tails can be 
deduced rather generally using very simple scaling arguments. The 
central idea is that large defects velocities are associated with the 
vanishing of defects at small length scales. Hence the tail gives 
information about the small-scale structure of the defect morphology. 
However, the velocity distribution is more cleanly defined and 
therefore more convenient to study theoretically (and in simulations) 
than the morphology itself. The physical arguments we shall use are 
very general, and can be applied to systems with both point-like and 
extended defects, with or without conservation of the order parameter.

Consider, for example, systems described by a scalar order parameter. 
This class includes the `standard' phase-ordering systems such as binary 
alloys (order-disorder transitions or phase separation) and binary liquids. 
The configuration of the coarsening system is described by the locations 
of the interfaces, or domain walls, that separate regions occupied by 
the two ordered phases. We are interested in the velocity distribution 
of points on the interfaces. The high-velocity tail in the distribution 
arises as follows. At all stages of the coarsening process, small domains 
are being eliminated. When the domains are very small, the walls move 
very quickly. So the high-velocity tail is related to the density of 
small domains. This is the basic idea behind the scaling argument 
presented in section II. For the case of non-conserved scalar fields, 
an explicit calculation using a `gaussian closure' scheme (section III) 
gives an approximate form for the full velocity distribution, extending 
Mazenko's work \cite{Mazenko} on the velocity distribution 
of point defects in $O(n)$ models. The tail of the distribution agrees 
with the result derived from scaling arguments.

We shall show that an alternative (and {\em a priori} equally valid) way 
of using the gaussian closure results to compute the velocity distribution 
gives a gaussian tail rather than a power-law tail. The defect velocity 
distribution, therefore, is a useful discriminator between various 
computational schemes. It is reassuring that, in the limit of large 
spatial dimension $d$, when the gaussian closure approximation is believed 
to become exact (for nonconserved fields), both schemes yield the same 
results.

In section IV the scaling arguments are extended to vector fields, and 
the general result $p=2+(d+1-n)/(z-1)$ is obtained for the tail exponent. 
For point defects ($n=d$) this result agrees with that obtained 
by Mazenko \cite{Mazenko} using the gaussian closure approximation. 
On the basis of the results for scalar fields and vector fields with 
point defects, we conjecture [equation (\ref{conj})] a general form for 
the defect velocity distribution within the gaussian approximation. 

\section{Scaling Approach}
\subsection{Nonconserved Scalar Fields} 
A system described by a nonconserved scalar field coarsens by curvature 
driven growth \cite{Review}. The normal velocity of a point on an interface 
is proportional to the total curvature $K$. For a small circular (or 
spherical, in $d=3$) domain of radius $r$, $K=(d-1)/r$. The velocity 
of the interface is therefore $v = dr/dt \propto -1/r$. It follows that 
the time $t$ for a domain of initial radius $r$ to disappear scales as 
$t \propto r^2$. 

Let $n(r)dr$ be the number of domains per unit volume with radius 
between $r$ and $r+dr$. The assumed scaling of the domain morphology gives 
\begin{equation}
n(r) = L^{-(d+1)}\,f(r/L)\ ,
\label{scaling}
\end{equation}
where $L(t) \sim t^{1/2}$ is the characteristic length scale at time $t$ 
after the quench into the ordered phase. Integrating (\ref{scaling}) 
over $r$ gives of order one domain per scale volume $L^d$, as required  
by scaling. 

In a time interval $\Delta t$, domains with $r < (\Delta t)^{1/2}$ will 
disappear. From (\ref{scaling}) the number of such domains is
\begin{equation}
L^{-(d+1)}\int_0^{(\Delta t)^{1/2}}dr\,f(r/L) 
\sim L^{-(d+1)}\,(dL/dt)\,\Delta t\ .
\label{dis}
\end{equation}
The right-hand side of (\ref{dis}) follows from the fact that the total 
number of domains per unit volume scales as $L^{-d}$. The requirement 
that the left-hand side of (\ref{dis}) be linear in $\Delta t$ for 
$\Delta t \to 0$ implies the property $f(x) \sim x$, for $x \to 0$, for 
the scaling function $f(x)$. Inserting this form in (\ref{dis}), one 
recovers $L(t) \sim t^{1/2}$ as expected.  

We have shown that $n(r)dr \sim rdr/L^{d+2}$ for $r \ll L$. The interfacial 
area per unit volume associated with these small domains is 
\begin{equation}
A(r)dr \sim r^{d-1}\,n(r)\,dr \sim r^d\,dr/L^{d+2}\ .
\label{area}
\end{equation}
Normalizing by the total interfacial area per unit volume, $L^{-1}$, 
gives the area-weighted probability for interfacial radius of curvature 
between $r$ and $r+dr$, i.e.\ 
\begin{equation}
P_r(r)dr \sim r^d\,dr/L^{d+1}\ , \ \ \ r \ll L\ .
\label{prob}
\end{equation}
Since the velocity associated with radius-of-curvature $r$ is $v \sim 1/r$, 
the interfacial velocity distribution $P_v(v)$ is obtained from 
$P_v(v)=P_r(r)|dr/dv|$, giving the power-law tail
\begin{equation}
P_v(v) \sim \frac{1}{v(vL)^{d+1}}\ .
\label{tail}
\end{equation}
This, together with its generalizations below, is our main result.

\subsection{Conserved Scalar Fields}
For conserved scalar fields, coarsening proceeds through the 
Lifshitz-Slyozov-Wagner (LSW) evaporation-condensation mechanism 
\cite{LSW}, by which large domains (or, more generally, regions 
with low interfacial curvature) 
grow at the expense of small domains (regions of high curvature). This 
leads to the LSW growth law $L(t) \sim t^{1/3}$. Large domain-wall 
velocities are associated, as in the non-conserved case, with the 
disappearance of small domains. For the conserved case, however, the 
relation between velocity and radius is $v(r) \sim 1/r^2$ 
\cite{Review,LSW}. Therefore, 
the time taken for a domain of size $r \ll L(t)$ to evaporate is of 
order $r^3$. In a time interval $\Delta t$, domains of size 
$r < (\Delta t)^{1/3}$ will disappear. Using (\ref{scaling}), the number 
of such domains is 
\begin{equation}
L^{-(d+1)}\int_0^{(\Delta t)^{1/3}}dr\,f(r/L) 
\sim L^{-(d+1)}\,(dL/dt)\,\Delta t\ ,
\label{disB}
\end{equation}
in complete analogy to (\ref{dis}). The requirement that the 
left-hand-side be proportional to $\Delta t$ forces the domain-size 
distribution function $f(x)$ to have the small-$x$ form $f(x) \sim x^2$ 
for conserved scalar fields. Inserting this form into (\ref{disB}) one 
recovers $L(t) \sim t^{1/3}$ as expected. 

The interfacial area per unit volume associated with these small domains 
is given by $A(r)dr \sim r^{d-1}n(r)dr$, as in (\ref{area}). 
Using $n(r)dr \sim r^2dr/L^{d+3}$ for $r \ll L$, which follows from 
$f(x) \sim x^2$, and normalizing by the interfacial area per unit volume, 
$L^{-1}$, gives the area-weighted probability for interfacial radius of 
curvature between $r$ and $r+dr$:
\begin{equation}
P_r(r)dr \sim r^{d+1}\,dr/L^{d+2}\ , \ \ \ r \ll L\ ,
\label{probB}
\end{equation}
instead of (\ref{prob}).  The final step is to use the relation 
$v \sim 1/r^2$ to deduce the velocity distribution $P_v(v)$ from 
$P_v(v)= P_r(r)|dr/dv|$. From (\ref{probB}) one infers the power-law 
tail 
\begin{equation}
P_v(v) \sim \frac{1}{v(vL^2)^{(d+2)/2}}
\label{tailB}
\end{equation}
for conserved scalar fields. 

The scaling approach adopted for scalar fields can be readily generalized 
to vector fields. Before doing so, however, we will consider an analytical 
approach, based on a gaussian closure scheme, for computing the full 
distribution $P_v(v)$ (i.e.\ not just the tail) for the nonconserved 
scalar case. This is instructive as it provides, in conjunction with the 
scaling result $(\ref{tail})$ for the tail, a way of discriminating 
between two ({\em a priori} equally valid) ways of using the gaussian 
closure results in the calculation of $P_v$.

\section{Calculation of $P_v(v)$ using Gaussian Closure}
The idea underlying all gaussian closure schemes is the introduction 
of a `smooth' auxiliary field $m({\bf x},t)$ whose zeros define the 
interfaces, i.e.\ $m$ has its zeros at the the same points as the 
order-parameter field $\phi({\bf x},t)$. Whereas $\phi$ is essentially 
constant within domains, but varies rapidly within domain walls, $m$ 
is `smooth', i.e.\ it varies only on the larger scale $L(t)$. At this 
stage, however, it is not necessary to specify precisely how $m$ is 
defined. 

An expression for the interface velocity in terms of $m$ can be obtained 
by noting that the rate of change of $m$ in a frame moving with the 
interface is zero, i.e.\ $dm/dt = 0 = \partial m/\partial t + 
{\bf v}\cdot\nabla m$. Since $\nabla m$ is normal to the interface, 
${\bf v}\cdot\nabla m = v_n|\nabla m|$, where $v_n$ is the normal 
velocity of the interface. This gives
\begin{equation}
v_n({\bf x}) = -\partial_t m/|\nabla m|\ .
\label{v}
\end{equation}
Formally this equation defines a `velocity' at every point -- the 
velocity of the surface of constant $m$ defined at that point. To 
find the distribution of interface velocities, we have to project 
onto the interfaces:
\begin{equation}
P_v(v) = \frac{\langle \delta(v - v_n({\bf x}))
\rho({\bf x})\rangle}{\langle \rho({\bf x})\rangle}\ ,
\label{probm}
\end{equation}
where $\rho({\bf x})$ is the areal density of interface, given by
\begin{equation}
\rho({\bf x}) = \delta(m({\bf x}))\,|\nabla m|\ .
\label{rho}
\end{equation}
Integrating $\rho({\bf x})$ over any volume of space gives the interfacial 
area in that volume. 

Equations (\ref{v} -- \ref{rho}) are exact, but in order to make further 
progress one needs to know the joint distribution function for $m$, 
$\partial_t m$, and the components $\partial_i m$ of $\nabla m$, at a 
given point in space. In all gaussian closure schemes, the field 
$m({\bf x},t)$ is assumed to be gaussian. This approximation works 
reasonably well in practice for nonconserved fields \cite{Review}. 
Using it, the required distribution function is expressible in terms 
of the inverse of the covariance matrix of the variables $m$, 
$\dot{m} \equiv \partial_t m$, and $\partial_i m$ ($i=1,\ldots,d$). 
Clearly $\langle m \partial_i m\rangle$ 
and $\langle \dot{m}\partial_i m \rangle$ both vanish due to 
translational invariance. Similarly, $\langle \partial_i m 
\partial_j m \rangle$ vanishes for $i\ne j$ due to the assumed isotropy 
of the system. It follows that the only non-zero elements of the covariance 
matrix are $S(t) = \langle m^2 \rangle $, $\dot{S}/2 = 
\langle m \dot{m} \rangle$, $T(t) = \langle \dot{m}^2 \rangle$ and 
$U(t) = \langle (\partial_i m)^2 \rangle$. Note that the $\nabla m$ 
sector is decoupled from the $(m,\dot{m})$ sector. The required probability 
distribution has, therefore, the product form 
$$
P(m,\dot{m},\nabla m) = \frac{1}{(2\pi U)^{d/2}}\exp\left( 
-\frac{(\nabla m)^2}{2U}\right)
$$
\begin{equation}
\times \frac{1}{2\pi D}\exp\left(
-\frac{1}{2D}(Tm^2 - \dot{S} m \dot{m} + S \dot{m}^2)\right),
\label{jointprob}
\end{equation}
where
\begin{equation}
D = ST - \dot{S}^2/4
\end{equation}
is the determinant of the covariance submatrix in the $(m,\dot{m})$ sector. 

From (\ref{probm}) -- (\ref{rho}) the velocity distribution is give by 
\begin{equation}
P_v(v) = \frac{\langle \delta(v + \dot{m}/|\nabla m|)\,\delta(m)\,
|\nabla(m)|\rangle}{\langle \delta(m)|\nabla(m)|\rangle}\ .
\end{equation}
Evaluating the averages using the distribution (\ref{jointprob}) gives
\begin{equation}
P_v(v) = \frac{1}{\sqrt{\pi}}\,\frac{\Gamma(\frac{d+2}{2})}
{\Gamma(\frac{d+1}{2})}\,
\frac{1}{\bar{v}}\,\frac{1}{(1+v^2/\bar{v}^2)^{(d+2)/2}}\ ,
\label{dist}
\end{equation}
where the characteristic velocity $\bar{v}$ is given by 
\begin{equation}
\bar{v} = \sqrt{\frac{D}{SU}} = \sqrt{\frac{\langle m^2 \rangle\,
\langle \dot{m}^2 \rangle - \langle m\dot{m}\rangle^2}
{\langle m^2 \rangle\,\langle (\nabla m)^2\rangle /d}}\ .
\label{v_0}
\end{equation}
Note that the distribution (\ref{dist}) is normalized on the interval 
$(-\infty,\infty)$.

The $v^{-(d+2)}$ tail predicted by (\ref{dist}) agrees with the 
scaling arguments for non-conserved fields presented in the previous 
section. This agreement is a nontrivial result, since up to now we 
have made no assumptions about the nature of the dynamics! The only 
assumption used is that the field $m$ is gaussian, which is evidently 
qualitatively correct for nonconserved fields but not for conserved 
fields, where the power-law predicted using (\ref{dist}) disagrees 
with the scaling result (\ref{tailB}). We conclude that no simple 
gaussian auxiliary-field approach captures the physics of the conserved 
model. Similar conclusions have been drawn in another context \cite{OSY}.  

The explicit value of $\bar{v}$ in (\ref{v_0}) depends on the dynamics  
of the field $m$. The simplest approximation is that of Ohta, Jasnow 
and Kawasaki (OJK) \cite{OJK}, in which $m$ obeys the diffusion equation, 
$\partial_t m = D_0\nabla^2 m$, with $D_0=(d-1)/d$. The initial condition is
$\langle m({\bf x},0) m({\bf x}',0) \rangle = 
\Delta \delta({\bf x}-{\bf x}')$. With this choice one obtains 
$\bar{v} = [(d-1)/2t]^{1/2}$. Other approximations, such as that of 
Mazenko \cite{GM1}, give the same $t^{-1/2}$ behaviour, but with a 
different prefactor. The $t^{-1/2}$ dependence follows quite generally 
from the scaling relation $\bar{v} \sim dL/dt$ and the usual result 
$L \sim t^{1/2}$ for nonconserved scalar fields. 

Before concluding this section, it is instructive to briefly discuss  
an alternative approach \cite{Bray93} to implementing the gaussian 
assumption for the field $m$. Consider a given mapping $\phi(m)$ between 
the original field $\phi$ and the auxiliary field $m$. The spatial 
variation of $\phi$ across an interface is asymptotically identical 
(in the scaling regime) to that of an equilibrium interface. It follows 
that, {\em at an interface}, $|\nabla \phi|$ is a given constant, 
determined by the equilibrium interface profile, and therefore 
$|\nabla m|$ is a constant, which we shall call $a$. 

This discussion shows that the gaussian approximation cannot be exact 
because, for a gaussian field $m$, $|\nabla m|$ fluctuates with position 
on the interface. An exception is the limit $d \to \infty$, when 
$|\nabla m|$ approaches the limit 
$[\sum_{i=1}^d \langle(\partial m/\partial x_i)^2\rangle]^{1/2}$. 
We shall return to this limit below. 

Replacing $|\nabla m|$ by $a$ in (\ref{v}) and (\ref{rho}) leads to the 
probability distribution 
\begin{equation}
P_v(v) = \frac{\langle \delta\left(v + \frac{\dot{m}}{a}\right)\,
\delta(m)\rangle}{\langle \delta(m) \rangle} = 
\left(\frac{Sa^2}{2\pi D}\right)^{1/2}\,
\exp\left(-\frac{Sa^2}{2D}\,v^2\right)\ .
\label{fixedwidth}
\end{equation}

This new approach, therefore, fails to reproduce the power-law tail 
in $P_v(v)$ predicted using the general arguments presented in 
section II, but gives instead a gaussian distribution for all $d$. 
While both approaches would seem to be equally valid {\em a priori}, 
the calculation of $P_v(v)$ discriminates clearly in favor of the 
first approach of treating the field $m$ consistently as gaussian 
throughout. In the limit $d \to \infty$, however, both approaches 
agree, since (\ref{dist}) has, for large $d$, the limiting gaussian form 
\begin{equation}
P_v(v) = (t/\pi)^{1/2}\,\exp(-tv^2)\ ,  
\label{distBH}
\end{equation}
using $\bar{v} \to (d/2t)^{1/2}$ from the OJK theory. 

A concrete realization of the new scheme requires a model which ensures 
that the interface thickness is time-independent. Although it is difficult 
to enforce the constraint $|\nabla m| = a$ at interfaces, models in 
which the looser condition $\langle (\nabla m)^2 \rangle = a^2$ is imposed 
can be devised. The simplest such model is that proposed by Bray and Humayun 
\cite{Review,BH92}, equivalent to the Oono-Puri extension \cite{OP} 
of the OJK model. Within this model, the field is still strictly gaussian, 
so $P_v(v)$ still has the form (\ref{dist}). This model allows, however, 
a quantitative comparison of (\ref{fixedwidth}) and (\ref{dist}) for 
$d \to \infty$.  One finds \cite{Review,BH92} $S=4t/d$, $T=(d+2)/2dt$, 
and $U=1/d$ (corresponding to $a=1$), and hence 
$D=2/d$. Putting these results into either (\ref{dist}), with $d \to \infty$, 
or into (\ref{fixedwidth}), with $a=1$, reproduces (\ref{distBH}).

\section{Vector Fields}
The scaling arguments of section II can be generalized rather simply to  
$n$-component vector fields in spatial dimension $d \ge n$ (the latter 
condition being necessary for the existence of localized defects). 
The quantitative (though approximate) methods of section III are, for 
technical reasons, less straightforward to extend to the general case. 
Mazenko \cite{Mazenko} has recently given the result for nonconserved 
fields with point defects ($n=d$). 

The scaling argument can be presented in a rather general way. If the 
dynamical exponent is $z$, i.e.\ $L(t) \sim t^{1/z}$, it follows that the 
characteristic velocity scales as $dL/dt \sim L^{-(z-1)}$. The next 
step is to assume that a small defect structure [a small domain ($n=1$), 
a small vortex loop ($n=2$, $d=3$) or vortex-antivortex pair ($n=2=d$), 
etc.], of size much less than $L(t)$, collapses under its own internal 
forces at a rate consistent with scaling. That is, the scale $r$ of the 
structure evolves as $dr/dt \sim -r^{-(z-1)}$ for $r \ll L(t)$. The 
collapse time of such a structure will then scale as $r^z$, and  
structures with $r < (\Delta t)^{1/z}$ will disappear in a time interval 
$\Delta t$. This scenario agrees precisely with what we found explicitly 
for scalar fields in section II, where $z=2$ (3) for nonconserved 
(conserved) fields. 
The case $n=2$ has been discussed by Rutenberg and Bray \cite{RB}. 

The number of defect structures per unit volume with size between $r$ and  
$r+dr$ is again given by  the scaling form (\ref{scaling}). Requiring 
that the number of structures which disappear in interval $\Delta t$ 
be linear in $\Delta t$ forces the scaling function $f(x)$ in 
(\ref{scaling}) have the small-argument form $f(x) \sim x^{z-1}$, giving 
$n(r) \sim r^{z-1}/L^{d+z}$ for $r \ll L$. The core volume of a structure 
of size $r$ scales as $r^{d-n}$, while the total core volume per unit 
volume of space scales as $L^{-n}$. The core-volume weighted probability 
for defect structure size between $r$ and $r+dr$ is, therefore,
\begin{equation}
P_r(r)\,dr \sim r^{z-1+d-n}dr/L^{d+z-n}\ ,\ \ \ r \ll L \ ,
\end{equation}
a generalization of (\ref{prob}) and (\ref{probB}).   
Using $v \sim r^{-(z-1)}$ leads to the power-law tail
\begin{equation}
P_v(v) \sim \frac{1}{v(vL^{z-1})^{p-1}}\ , 
\label{probgen}
\end{equation}
with tail-exponent
\begin{equation}
p = 2 +(d+1-n)/(z-1)\ .
\label{p}
\end{equation}
This result generalizes (\ref{tail}) and (\ref{tailB}), to which it reduces 
for $n=1$ and $z=2,3$ respectively.

We can compare this result with the analytical result for $P_v(v)$ obtained 
using the gaussian closure approximation for nonconserved fields ($z=2$) 
with point defects ($n=d$)\cite{Mazenko}. For this case (\ref{p}) becomes 
$p=3$, i.e.\ $P_v(v) \sim 1/L^2v^3$. Note that $P_v(v)$ is the probability 
distribution for the {\em magnitude} of the velocity. The gaussian closure 
calculation gives an expression for the probability distribution 
$P(\vec{v})$ of the {\em vector} velocity, in the form \cite{Mazenko} 
\begin{equation}
P(\vec{v}) = \frac{\Gamma(\frac{n+2}{2})}{(\pi \bar{v}^2)^{n/2}}\, 
\frac{1}{(1+v^2/\bar{v}^2)^{(n+2)/2}}\ .
\end{equation}
The normalization of this distribution is $\int d^nv\, P(\vec{v}) =1$, 
whereas $P_v(v)$ is normalized as $\int_0^\infty dv\,P_v(v)=1$. 
The relation between these two distributions is simply 
$P_v(v) = [2\pi^{n/2}/\Gamma(n/2)]v^{n-1}P(v)$, i.e.\ 
\begin{equation}
P_v(v) = \frac{n}{\bar{v}}\,\frac{(v/\bar{v})^{n-1}}
{(1+v^2/\bar{v}^2)^{(n+2)/2}}\ .
\label{distvec}
\end{equation}
The power-law tail is $P_v(v) \sim n\bar{v}^2/v^3$, which agrees with the 
scaling result (\ref{p}) (for the case $n=d$, $z=2$). 

\section{Conclusion}
The velocity distribution of topological defects in a phase-ordering 
system has been discussed. A simple scaling argument, introduced in the 
context of domain walls in section II, and generalized to vector fields 
in section IV, predicts a power-law tail at large velocity. The 
tail exponent $p$ is given by the general result (\ref{p}), where $z$ is 
the dynamical exponent that describes the coarsening dynamics, via  
$L(t) \sim t^{1/z}$. 

For the special case of non-conserved fields, approximate analytical 
calculations, based on a gaussian closure scheme, have been performed 
for scalar fields (section III) and for vector fields with point defects 
\cite{Mazenko}. In both cases the approximate result exhibits the  
power-law tail predicted by the scaling arguments. Comparison of 
equations (\ref{dist}) and (\ref{distvec}) suggests the following 
conjecture for general nonconserved vector fields, within the gaussian 
approximation: 
\begin{equation}
P_v(v) = \frac{2\Gamma(\frac{d+2}{2})}{\Gamma(\frac{n}{2})
\Gamma(\frac{d-n+2}{2})}\,\frac{1}{\bar{v}}\,\frac{(v/\bar{v})^{n-1}}
{(1+v^2/\bar{v}^2)^{(d+2)/2}}\ .
\label{conj}
\end{equation}
This result reduces to (\ref{dist}) for $n=1$ \cite{Note}, to 
(\ref{distvec}) for $n=d$, and gives rise to the power-law tail 
$P_v(v) \sim v^{-p}$ with $p=d+3-n$, in agreement with the scaling 
prediction (\ref{p}) for $z=2$. 

As a final comment we note that for certain systems the growth law for 
$L(t)$ is expected to contain logarithmic corrections to a simple power 
law. Two examples are the nonconserved $n=2$ model for $d=2$, where 
one expects \cite{RB,YPKH} $L(t) \sim (t/\ln t)^{1/2}$, and the 
conserved $n=2$ model for $d \ge 3$, where \cite{RB,BR} 
$L(t) \sim (t \ln t)^{1/4}$. A simple scaling approach of the type 
used in sections II and IV cannot be applied naively, due to the 
appearance of a new length scale -- the defect core size -- which 
enters as a short-scale cut-off in the logarithms. We would expect, 
however, that any resulting modifications of the final result would 
be limited to possible logarithmic corrections to the power-law tail 
in $P_v(v)$. The dominant power-law part will be obtained by inserting 
the appropriate value of $z$ (2 or 4 respectively in the cases discussed 
above) in (\ref{p}). 

\begin{center}
\begin{small}
{\bf ACKNOWLEDGMENTS}
\end{small}
\end{center}
It is a pleasure to thank Andrew Rutenberg for stimulating discussions 
on defect dynamics over several years. The seeds of the scaling ideas 
developed in sections II and IV were, I am sure, planted during these 
early discussions. I thank Gene Mazenko for a useful correspondence. 

This work was supported by EPSRC grant GR/K53208.

\end{multicols}

\begin{references}
\bibitem{Note} The factor 2 discrepancy between (\ref{dist}) and 
(\ref{conj}) with $n=1$ simply reflects the different intervals on which 
the two distributions are normalized -- the former on $(-\infty,\infty)$, 
the latter on $(0,\infty)$.
\end{references}
\end{document}